\documentclass[12pt,a4paper]{article}
\usepackage[T2A]{fontenc}

\def\d{\displaystyle}
\begin{document}
\centerline {\bf INFLUENCE OF GRAVITATIONAL LENSING}

\centerline {\bf ON THE DETERMINATION OF THE LUMINOSITY}

\centerline {\bf FUNCTION OF BACKGROUND OBJECTS}
\vskip 0.4 cm

\centerline {Yu. L. Bukhmastova, bukh@astro.spbu.ru}
\vskip 0.7cm

The distribution function of quasars with respect to apparent brightness is given, found under the assumption
that quasars are, at least partially, the gravitationally enhanced images of the active nuclei of distant
galaxies. A Schechter law and a two-power law for the luminosity function of the sources are used, as well as
a probabilistic law of image enhancement for various models of gravitational lenses. To find the theoretical
distribution function of quasars with respect to apparent brightness we use a theorem on the probability
density of a product of random quantities. It is shown that the slope of this function ranges from $-1$ to 
$-2$ for faint quasars, like that for ordinary galaxies. In the case of bright quasars, the slope of the apparent
brightness distribution function is determined mainly by the lensing effect and has a lower limit of $-3.$ The
good agreement between theory and observations suggests that statistically quasars are gravitationally
enhanced images of the active nuclei of distant galaxies. If the initial assumptions are correct, then the
luminosity function of galaxies and the apparent brightness function for quasars are not independent but are
related by means of the differential lensing probability.
\vskip 0.3 cm

\vskip 0.5 cm

{\bf 1. Introduction}
\vskip 0.2 cm 
Light rays from a distant source, passing by a massive object or gravitational lens, undergo deviation. This can
increase the luminous area of the source. This means that the brightness of a distant source can increase by tens or
hundreds of times due to the gravitational focussing effect. This can mean that the distribution function of sources with
respect to apparent brightness, obtained on the basis of an analysis of various observations, will erroneously be taken
as the luminosity function of the sources. The influence of gravitational lensing effects must thus be taken into account
in an analysis of the luminosity function of sources.
The present work is a logical extension of [1, 2], in which the possible influence of gravitational lensing effects
on the appearance of close quasar-galaxy pairs was discussed. In the present work we suggest a method enabling one
to obtain the analytical form of the apparent brightness distribution function of sources (Sec. 4) with allowance for the
influence of gravitational lensing (Sec. 3). It is proposed that the source luminosity function be analyzed in the form
of a Schechter function or a two-power law (Sec. 2).

It is proposed to apply this method to the analysis of the quasar luminosity function as an example (Sec. 5). As
the basis for the arguments we take the Barnothy-Tyson hypothesis [3-6], according to which at least some quasars are
gravitationally enhanced images of compact massive objects lying in the central regions of distant galaxies. Such compact
objects may be black holes with masses on the order of $10^6-10^{9.5} M_{\odot}$, [7-11], which, it is generally accepted, lie in the
central regions of most galaxies. If the surface density of the number of normal galaxies with a stellar magnitude down
to $28^m-29^m$ is on the order of $10^6 gal./deg^2,$ which follows from observations on the Hubble space telescope, then even
with a low probability of $10^{-4}-10^{-6}$ for brightness enhancement by $3^m-5^m,$ one can expect from 1 to 100 quasars per square
degree by preliminary rough estimates. At present it is difficult to estimate the stellar magnitudes of the presumed quasars,
since for the lensed sources one takes not the galaxies themselves but the compact massive objects in their central regions.
It was pointed out in [12] that the luminosities of the nuclei of 42 Seyfert galaxies correlate with the luminosities of the
parent galaxies. It was pointed out in [13] that the masses of the compact massive objects are in a strict percentage ratio
to the masses of the galaxies as a whole and comprise $\sim 0.2\%$ of the total mass of an elliptical parent galaxy or of the
mass of the bulge of a spiral galaxy. All this suggests that the luminosities of the nuclei and the luminosities of the parent
galaxies are in some percentage ratio. Hence it follows that the luminosity function of the compact massive objects or
sources and the luminosity function of galaxies are identical in general form, and hence the luminosity function of the
sources can be analyzed in the form of a Schechter law or a two-power law.
In Sec. 5 of the present paper we show that in the case of faint quasars the observational data agree well both
with the Schechter form of the source luminosity function and with the two-power law. The slope of the function (the
exponent in the case of a two-power law) is from $-1$ to $-2,$ just as for galaxies.
A new result is that the theoretical slope of the function for bright quasars has a lower limit of $-3$ rather than
$-2,$ as stated in [14]. This theoretical limit for bright quasars is determined mainly by the lensing effect, in contrast to
faint quasars, for which lensing has almost no effect on the slope.
In Sec. 5 an attempt is made to reconcile the theoretical result with observational data. The main conclusions
of the work are briefly formulated in Sec. 6.
It should be noted that the spectra of lensed quasars are a separate problem, which is not treated in the present
work.

\vskip 0.3 cm
{\bf 2. Luminosity Function of Background Sources}
\vskip 0.2 cm

We assume that the luminosity function of the sources can be represented in the form of the Schechter function,
which is valid for galaxies, or a two-power law, which is applied in particular for the description of galaxies and quasars
[14-19].
The general form of the Schechter function is

$$p_L(y)=n_0\cdot y^{\alpha}e^{-y},\eqno(1) $$
$a<y<b,\hskip 10pt$ $n_0=\frac{\d 1}{\d{\Gamma(1+\alpha,a)}-\d{\Gamma(1+\alpha,b)}},\hskip 10pt$ 
$-2\le \alpha\le -1,$
 where $y=L/L_0, \hskip 10pt L_0-$  being the characteristic luminosity of the sources.
Here $a$ and $b$ are the lower and upper limits of the relative luminosity and $\Gamma(\alpha,x)$ is a partial gamma function.
The luminosity function of sources in the form of a two-power law is
$$p_L(y)=\cases{c_1y^{\alpha},\hskip 10pt a<y\leq k,\cr c_2y^{\beta},\hskip 10pt k<y<b.}\eqno(2)$$
Here $a$ and $b$ are the lower and upper limits of the relative luminosity, $k$ is the boundary of the break in the two-power
law, and $c1$ and $c2$ are constants, and $\alpha$ and $\beta$ are exponents determining the slope of the luminosity 
function for faint and bright sources, respectively.

\vskip 0.3 cm
{\bf 3. Differential Lensing Probability}
\vskip 0.2 cm

We assume that all sources lie at some average fixed distance (corresponding to a redshift $z\sim 1$) and we consider
the case of considerable brightness enhancements $(\mu > 1).$ For various models of gravitational lenses [1, 14], the
differential probability that a source with a fixed redshift will be enhanced by a factor of $\mu$ is defined as

$$p_A(\mu)=\frac{\nu-1}{\mu^{\nu}},\hskip 10pt \mu>1, \nu>1.\eqno(3)$$
The value of $\nu$ is close to 3.

\vskip 0.3 cm
{\bf 4. Distribution Function of 

Apparent Brightness of Lensed Sources}
\vskip 0.2 cm

Having the source luminosity function (1) or (2) and the probability (3) that sources will be enhanced by a factor
of $\mu$, we can determine the apparent brightness function for lensed sources. For this we must use a theorem on the
probability density of a product of independent random quantities [20]. We define the apparent luminosity $l$ of a source
as the product of the source's luminosity $L$ times the amplification ratio $A.$ We thus have $l = A*L.$ We consider the
probability distribution densities of three random quantities, $p_l(x), p_A(m)$ and $p_L(y).$ We introduce the variable of
integration $u = x/\mu.$ The probability density for the distribution of lensed sources with respect to apparent brightness is
defined as
$$p_l(x)=\int_{0}^{\infty} p_L(u)\cdot p_A\Bigl(\frac{x}{u}\Bigr)\frac{du}{u}.\eqno(4)$$
With allowance for (1) and (3), we have
$$p_l(x)=n_0\frac{\nu-1}{x^\nu}\biggl[\Gamma(\nu+\alpha,a)-\Gamma(\nu+\alpha,\min\{x,b\})\biggr].\eqno(5)$$
The slope of this function is defined as
 $\beta(x)=\d{-\frac{ {d\ln} p_l(x)}{ {d\ln}x}}.$
With allowance for (5), we have
$$\beta(x)=-\cases{\d{\nu-\frac{x^{\nu+\alpha}e^{-x}}{\Gamma(\nu+\alpha)-\Gamma(\nu+\alpha,x)}},\hskip 10pt a<x<b,\cr
\nu,\hskip 10pt x>b.}\eqno(6)$$
Using (2) instead of (1), with allowance for (3) and (4) we obtain the following expressions for the apparent
brightness function for lensed sources:

 in the case of faint sources
$$p_l(x)\sim c_1c_2\Biggl(\frac{x^{\alpha}}{\nu+\alpha}-\frac{x^{-\nu}a^{\nu+\alpha}}{\nu+\alpha}\Biggr),\eqno(7)$$

in the case of bright sources
 $$p_l(x)\sim c_2c_3\Biggl(\frac{x^{\beta}}{\nu+\beta}-\frac{x^{-\nu}k^{\nu+\beta}}{\nu+\beta}\Biggr).\eqno(8)$$
\vskip 0.3cm

{\bf 5. Apparent Luminosity of Quasars}
\vskip 0.2 cm

Many authors have said that quasars may be objects enhanced by gravi-tational lensing. It was suggested in [14],
in particular, that the brightness of quasars may be enhanced by gravitational lenses closer to the observer. This means
that quasars may actually be fainter objects than they seem. The authors of [14] considered the luminosity function of
quasar sources in the form of a two-power law and used a differential lensing probability function to obtain the apparent
luminosity function, which turned out to be flatter than the observed function. They obtained a theoretical lower limit
of $-2$ on the exponent of the luminosity function of bright quasars, characterizing the slope of that function.
The quasar luminosity function is often represented in the form of a two-power law. On the basis of observational
data on more than 1000 quasars with $0.1 < z < 3.3$ [16] and some 200 quasars with $2.0 < z < 4.5$ [17], it was concluded
in [18] that the luminosity function for faint quasars is  $\Phi\sim L^{-1.7},$  while that for bright quasars is
  $\Phi\sim L^{-3.6}.$  The exponents
 $\alpha=-1.7$ and  $\beta=-3.6$ determine the slopes of the luminosity functions for faint and bright quasars, respectively. Values
of $-1.96<\alpha<-1.55$ and  $-3.75<\beta<-3.45$ are given in the 2dF QSO survey [19]. According to the observational data
of [21], $\alpha=-1.4,\hskip 10pt \beta=-2.6.$ There is thus a spread in the values of the exponents determined from different
observations: $-2<\alpha<-1,\hskip 10pt -3.75<\beta<-2.6.$ 

Let us assume that quasars are the enhanced active nuclei of distant galaxies. The luminosity function of the
sources then corresponds to the luminosity function of active galactic nuclei. Let us consider the Schechter form (1) of
the luminosity function of galactic nuclei. For galaxies we have $-2\le\alpha\le -1$, and for determinacy we take  $\alpha=-1.$ We
also take $\nu=3$ in Eq. (3). In this case, the distribution function of quasars with respect to apparent brightness with
allowance for (5) has the form
$$p_l(x)=\frac{2n_0}{x^3}\biggl[(a+1)e^{-a}-(x+1)e^{-x}\biggr], \eqno (9)$$
while the slope of the function is
$$ \beta(x)=-\cases{\d{\frac{3-x^2e^{-x}}{(a+1)e^{-a}-(x+1)e^{-x}}},\hskip 10pt a<x<b,\cr 3,\hskip 30pt x>b.}  \eqno(10)$$

In Fig. 1 we give the luminosity function for active galactic nuclei (a), the apparent brightness distribution function
for quasars (b), determined from Eq. (9), and the approximations for faint quasars  $(p_l(x)\sim x^{-1.4})$  and bright quasars
$(p_l(x)\sim x^{-2.6})$ in accordance with the observational data of [21]. It is seen from this figure that good agreement between
the theoretically obtained dependence $p_l(x)$ and the observational approximations is achieved for  $\alpha =-1.4, \beta=-2.6$ 
As follows from Eq. (10), the slope of the function for bright quasars has a lower limit of $-3,$ so agreement with other
observational data is achieved with some stretch.

 As an example, in Fig. 2 we give the apparent brightness distribution
function (9) for quasars with two observational approximations for faint quasars, with a slope  $\alpha=-1.7,$ and for bright
quasars, with   $\beta=-3.6$ [18].

In Fig. 3 we give graphs of the surface brightness function for quasars in the forms (7) and (8), as well as observed
approximations with  $\alpha=-1.7,\hskip 10 pt \beta=-3.2$ [22]. The adopted values of the parameters are
 $\nu=3,\hskip 0.2cm a=0.1,\hskip 0.2cm k=0.5, \hskip 0.2cm c_1c_2=0.6,\hskip 0.2cm c_2c_3=0.6,\hskip 0.2cm
\alpha=-1.8,\hskip 0.2cm \beta=-6.$

 The good visual agreement between the theoretical and observational surface
brightness distribution functions for quasars suggests that a correction evidently should be introduced into the observational
data with allowance for the threotically predicted lower limit of $-3$ for the exponent.
\vskip 0.3 cm

{\bf 6. Conclusion}
\vskip 0.2 cm

The gravitational lensing effect can affect the brightness of active nuclei-- compact sources in the central regions
of massive galaxies. Objects of increased brightness, which will have the observed properties of quasars, can appear in
this case. Using the differential lensing probability function, from the luminosity function for galactic nuclei the quasar
distribution function with respect to apparent brightness was obtained, which is often taken as the quasar luminosity
function. Two consequences were formulated on the basis of this assumption.
In the case of faint quasars, the lensing effect has little influence on the slope of the apparent brightness
distribution function, which in this case ranges from $-1$ to $-2,$ like the luminosity function for ordinary galaxies.
In the case of bright quasars the slope is determined mainly by the lensing effect and has a lower limit of
 $\beta=-3,$ in contrast to the result of [14], in which it is stated that the slope of the apparent brightness function for lensed
bright quasars cannot overstep the value $\beta=-2.$  

Our quasar distribution function (9) with respect to apparent brightness can have different power-law approximations.
This makes it possible to reconcile the theoretical result with various observational data, which is indirect
confir-mation of the lensed nature of the increased brightness of quasars. If such assumptions are correct, then the
luminosity functions of galaxies and quasars are not independent but are related by means of the differential lensing
probability.
Along with the quasar distribution function (5) with respect to surface brightness in general form, working
equations (7) and (8) were obtained for faint and bright quasars.
It should be noted that this method of allowing for the lensing effect in an analysis of the luminosity function
can be applied to any other objects.

The author thanks Yu. V. Baryshev for a useful discussion of this problem, as well as D. S. Bukhmastov for technical
assistance in preparing the publication.

\vskip 0.5 cm
{\bf REFERENCES}
\vskip 0.2 cm

1.{\it  Yu. V. Baryshev and Yu. L. Ezova ( Yu. L. Bukhmastova),}

 Astron. Zh., {\bf 74,} 497 (1997).
\vskip 0.2 cm

2.{\it  Yu. L. Bukhmastova,} Astron. Zh., {\bf 78,} 1 (2001).
\vskip 0.2 cm

3.{\it J. M. Barnothy,} Astron. J., {\bf 70,} 666 (1965).
\vskip 0.2 cm

4. {\it J. M. Barnothy,} Bull. Am. Astron. Soc., {\bf 6,} 212 (1974).
\vskip 0.2 cm

5. {\it J. A. Tyson,} Astrophys. J., {\bf 248,} L89 (1981).
\vskip 0.2 cm

6. {\it J. A. Tyson,} Astron. J., {\bf 96,} 1 (1988).
\vskip 0.2 cm

7.{\it  M. Bartusiak,} Astronomy, No. 6, 42 (1998).
\vskip 0.2 cm

8. {\it J. Kormendy,} astro-ph/0007400.
\vskip 0.2 cm

9.{\it  J. Kormendy,} astro-ph/0007401.
\vskip 0.2 cm

10.{\it  J. Kormendy and L. C. Ho,} astro-ph/0003268. 
\vskip 0.2 cm

11.{\it  L. C. Ho and J. Kormendy,} astro-ph/0003267.
\vskip 0.2 cm

12. {\it G. L. Granato, V. Zitelli, F. Bonoli, L. Danese,

 C. Bonoli, and F. Delpino,} Astrophys. J. Suppl. Ser., {\bf 89,} 35 (1993).
\vskip 0.2 cm

13. {\it  S. Nadis,} Astronomy, No. 2 (2001).
\vskip 0.2 cm

14. {\it P. Schneider, J. Ehlers, and E. E. Falko,} Gravitational Lenses,

 Springer-Verlag, New York (1992), p. 375.
\vskip 0.2 cm

15.{\it B. J. Boyle, T. Shanks, and B. A. Peterson,}

 Mon. Not. R. Astron. Soc., {\bf 235,} 935 (1988).
\vskip 0.2 cm

16. {\it F. D. A. Hartwick and D. Schade,}

 Annu. Rev. Astron. Astrophys., {\bf 28,} 437 (1990).
\vskip 0.2 cm

17. {\it S. J. Warren, P. C. Hewett, and P. S. Osmer,}

 Astrophys. J., {\bf 421,} 412 (1994).
\vskip 0.2 cm

18. {\it Y. C. Pei,} Astrophys. J., {\bf 438,} 623 (1995).
\vskip 0.2 cm

19. {\it B. J. Boyle, S. M. Croom, R. J. Smith, T. Shanks, P. J. Outram, 

F. Hoyle, L. Miller, and N. S. Loaring,} astro-ph/0103064.
\vskip 0.2 cm

20. {\it B. V. Gnedenko,} A Course in Probability Theory [in Russian], Nauka,

 Moscow (1988) [translated by B. Seckler: The Theory of Probability 

and the Elements of Statistics, 5th ed., Chelsea Publ., New York (1988)].
\vskip 0.2 cm

21. {\it H. L. Marshall, Y. Avni, A. Braccesi, J. P. Huchra, H. Tananbaum,

 G. Zamorani, and V. Zitelli,} Astrophys. J., {\bf 283,} 50 (1984).
\vskip 0.2 cm

22. {\it I. Kovner,} Astrophys. J., {\bf 341,} L1 (1989).

\end{document}